\documentclass[conference]{main}
\IEEEoverridecommandlockouts

\usepackage{cite}
\usepackage{amsmath,amssymb,amsfonts}
\usepackage{algorithmic}
\usepackage{graphicx}
\usepackage{textcomp}
\usepackage{xcolor}
\usepackage{multirow} 
\usepackage{amsmath}    
\usepackage{graphicx}   
\usepackage{ulem}       
\usepackage{graphicx}
\usepackage{subfigure}
\usepackage{hyperref}

\def\BibTeX{{\rm B\kern-.05em{\sc i\kern-.025em b}\kern-.08em
    T\kern-.1667em\lower.7ex\hbox{E}\kern-.125emX}}   
\begin{document}
\title{Chinese-LiPS: A Chinese Audio-Visual Speech Recognition Dataset with Lip-Reading and Presentation Slides \\
\thanks{$^{\dag}$ These authors contributed equally to this work.}
\thanks{$^{*}$ Corresponding author. This work has been supported by the National Key R\&D Program of China (Grant No.2022ZD0116307) and NSF China (Grant No.62271270).}
}

\author{\IEEEauthorblockN{Jinghua Zhao$^{1,\dag}$, Yuhang Jia$^{1,\dag}$, Shiyao Wang$^{1}$, Jiaming Zhou$^{1}$, Hui Wang$^{1}$, and Yong Qin$^{1,*}$}
\IEEEauthorblockA{
    $^{1}$College of Computer Science, Nankai University, Tianjin, China\\
    \textbf{Correspondence}: zhaojinghua\_hlt, 2120240729@mail.nankai.edu.cn, qinyong@nankai.edu.cn
    }
}
\maketitle

\begin{abstract}
Incorporating visual modalities to assist Automatic Speech Recognition (ASR) tasks has led to significant improvements. However, existing Audio-Visual Speech Recognition (AVSR) datasets and methods typically rely solely on lip-reading information or speaking contextual video, neglecting the potential of combining these different valuable visual cues within the speaking context. In this paper, we release a multimodal Chinese AVSR dataset, Chinese-LiPS, comprising 100 hours of speech, video, and corresponding manual transcription, with the visual modality encompassing both lip-reading information and the presentation slides used by the speaker. Based on Chinese-LiPS, we develop a simple yet effective pipeline, LiPS-AVSR, which leverages both lip-reading and presentation slide information as visual modalities for AVSR tasks. Experiments show that lip-reading and presentation slide information improve ASR performance by approximately 8\% and 25\%, respectively, with a combined performance improvement of about 35\%. The dataset is available at \url{https://kiri0824.github.io/Chinese-LiPS/}
\end{abstract}


\vspace{-4mm}
\section{Introduction}
\vspace{-0.8mm}

\label{sec:intro}
Automatic Speech Recognition (ASR) tasks often face performance challenges due to factors such as the speaker's tone, domain-specific terminology, and background noise. Incorporating information from modalities beyond speech can significantly improve the performance and robustness of ASR systems. Audio-Visual Speech Recognition (AVSR), which combines both audio and visual modalities, has shown significant potential in addressing these challenges. Existing AVSR approaches can be broadly categorized into two types: lip-reading-based methods \cite{ma2023auto, whisperflamingo, afouras2022deep, petridis2018audio, tseng2024av} and semantic visual contextual cues-based methods. The second type can be further divided into two subcategories: those that utilize speaking contextual videos \cite{peng2023prompting, gabeur2022avatar, wang2024slideavsr} and those that leverage a speaker’s presentation slides' textual information \cite{wang2024slidespeech, yang2024mala, yu2024lcb}. As these visual cues are not mutually independent in real-world scenarios. Integrating all potentially valuable visual information holds significant promise for further improving ASR performance.

However, there are two main challenges to achieving this integration of all potentially valuable visual information. First, there is a lack of high-quality datasets that incorporate all these types of visual information. SlideSpeech \cite{wang2024slidespeech} stands out as the first to leverage the presentation slides to assist in speech recognition. However, the videos in the SlideSpeech were automatically collected and annotated from wild media websites, raising concerns about the quality and consistency of both the presentation slides and their content. Additionally, SlideSpeech does not ensure the inclusion of accurate or well-captured lip movements, which limits its applicability for lip-reading-based ASR improvements. Moreover, there is currently no similar dataset available for Chinese. The second challenge lies in effectively integrating lip-reading with multiple types of semantic visual cues. Specifically, this involves determining how to acquire diverse and effective visual cues, how to fuse these cues with lip-reading features, and whether such integration can ultimately enhance performance.

To address the first challenge, we propose \textbf{Chinese-LiPS} (\textbf{Chinese} audio-visual speech recognition dataset with \textbf{Li}p-reading and \textbf{P}resentation \textbf{S}lides), a high-quality multimodal Chinese AVSR dataset comprising 100 hours of speech, video, and manually curated transcriptions. This dataset uniquely combines lip-reading videos with the speaker’s presentation slides, enabling a more comprehensive exploration of AVSR. Notably, the presentation slides in Chinese-LiPS are meticulously designed by domain experts, ensuring superior content quality, particularly in terms of the visual images’ quality and richness. To address the second challenge, we develop a basic but effective pipeline, LiPS-AVSR, which integrates both lip-reading and presentation slides for the AVSR task. This pipeline considers not only OCR-extracted (Optical Character Recognition) text but also visual cues from images and graphical content in the slides. Experimental results demonstrate that incorporating lip-reading and presentation slide information improves ASR performance by approximately 8\% and 25\%, respectively, with a combined gain of around 35\%. To summarize, our main contributions are as follows:

\begin{table*}[t!]
\caption{Comparison of Existing AVSR Datasets, Highlighting Lip-Reading and Contextual Information Availability}
\vspace{-3mm}
\begin{center}
\renewcommand{\arraystretch}{1.2}  
\begin{tabular}{c|ccc|cc|cc}
\hline
Dataset          & Speakers & Duration(h) & Language & Lip-reading Video& Contextual-cues(Slides/Video)& Year & Available\\ \hline
LRW\cite{chung2017lip} & 1000+ & - & English & \checkmark & - &  2018 & Y \\
LRS\cite{son2017lip} & - & - & English & \checkmark & - &  2017 & N \\
LRS2-BBC\cite{afouras2022deep} & - & 200 & English & \checkmark & - &  2018 & Y \\
LRS3-TED\cite{afouras2008lrs3} & - & 400 & English & \checkmark & - &  2018 & Y \\
LRW-1000\cite{yang2019lrw} & 2000+ & 57 & Chinese & \checkmark & - &  2018 & Y \\
CMLR\cite{zhao2019cascade} & - & - & Chinese & \checkmark & - &  2019 & Y \\
CN-Celeb-AV\cite{li2023cn} & 1,136 & 669 & Chinese & \checkmark & - &  2023 & Y \\
CN-CVS\cite{chen2023cn} & 2,557 & 300+ & Chinese & \checkmark & - &  2023 & Y \\
\hline
How2\cite{sanabria2018how2} & - & 2000 & Eng\&Por & - & \checkmark(instructional videos) &  2018 & Y \\
VisSpeech\cite{gabeur2022avatar} & - & 0.6 & English & - & \checkmark(Youtube videos) &  2022 & Y \\
SlideSpeech\cite{wang2024slidespeech} & - & 1000+ & English & - & \checkmark(presentation slides) & 2023 & Y \\
SlideAVSR\cite{wang2024slideavsr} & 220 & 36 & English & - & \checkmark(Youtube videos) &  2024 & Y \\
AVNS\cite{luo2024multi} & - & 30+ & English & - & \checkmark(background scenes) &  2024 & Y \\
\hline
3-Equations\cite{guan2024multi} & - & 25.2 & English & \checkmark & \checkmark(3-lines math formula) &  2024 & Y \\
Chinese-LiPS(ours) & 207 & 100 & Chinese & \checkmark & \checkmark(presentation slides) &  2024 & Y \\
\hline
\end{tabular}
\end{center}
\label{AVSR Datasets}
\vspace{-6mm}
\end{table*}
\vspace{-0.8mm}
\begin{itemize}
\item We introduce Chinese-LiPS, the first multimodal Chinese AVSR dataset that integrates both lip-reading information and presentation slides, comprising 100 hours of annotated speech, video, and transcriptions, with high-quality slides meticulously crafted by domain experts.
\item We develop a simple yet effective pipeline, LiPS-AVSR, which leverages lip-reading and presentation slides to enhance ASR performance by incorporating not only OCR-extracted text but also valuable visual cues from images and other graphical content.
\item We design experiments to validate the complementary roles of lip-reading and presentation slide information in enhancing ASR performance, highlighting the synergy of these visual modalities in addressing challenges such as ambiguous speech, and domain-specific terminology, demonstrating that integrating all potentially valuable visual information can further improve ASR performance.
\end{itemize}

\vspace{-1mm}
\section{Related Work}
\vspace{-0.8mm}

\subsection{Lip-Reading for AVSR}
\vspace{-0.8mm}

Lip-reading is one of the most accessible visual modalities for AVSR systems, capturing articulation-related details such as mouth movements and gestures. It is naturally synchronized with the audio modality in terms of timing, making it a valuable source of information. Consequently, most AVSR datasets emphasize lip-reading due to its practicality and ease of data acquisition.

For English, notable datasets include LRW \cite{chung2017lip}, which is constructed from television shows and captures a wide range of conversational interactions. The LRS series \cite{son2017lip, afouras2022deep, afouras2008lrs3} collects data from BBC broadcasts and TED talks, providing rich content from speeches and interviews. For Chinese, datasets such as CMLR \cite{zhao2019cascade} are gathered from news broadcasts, while CN-CVS \cite{chen2023cn} includes diverse real-world scenes like news programs and public speaking events, as outlined in Table~\ref{AVSR Datasets}.

Leveraging these datasets, several AVSR methods have been proposed. For instance, Auto-AVSR \cite{ma2023auto} and Whisper-Flamingo \cite{whisperflamingo} use feature fusion and cross-attention to integrate lip-reading and speech data. Additionally, AV-HuBERT \cite{shilearning} employs self-supervised learning to unify visual and audio information, significantly enhancing ASR performance.

\vspace{-1.2mm}
\subsection{Semantic Visual Contextual Cues for AVSR}
\vspace{-0.8mm}

Beyond lip-reading, semantic visual contextual cues play a crucial role in enhancing ASR tasks by providing supplementary information that complements the speech signal. These cues, which include elements like presentation slides, background scenes, and other visual context, help disambiguate spoken content and improve recognition accuracy.

For instance, in SlideSpeech \cite{wang2024slidespeech}, the textual and graphical content on slides has been shown to provide meaningful semantic context that enhances comprehension. In How2 \cite{sanabria2018how2} and AVNS \cite{luo2024multi}, background scenes have been demonstrated to offer insights into the environment where the speech takes place, further aiding speech recognition by incorporating contextual information beyond the speech signal. Additionally, datasets like VisSpeech \cite{gabeur2022avatar} and SlideAVSR \cite{wang2024slideavsr}, collected from general videos on video-sharing platforms, have also been found to effectively improve ASR performance.

Currently, most AVSR methods \cite{wang2024slidespeech, wang2024slideavsr} relying on semantic visual contextual cues primarily use OCR to detect text within contextual elements and leverage this textual information as keywords to enhance ASR performance. A few approaches \cite{peng2023prompting, guan2024multi} employ pre-trained multimodal models like CLIP  and DALL-E, to extract semantic information from these visual cues. Notably, \cite{guan2024multi} is the first to propose a dataset and method that combine speech, lip-reading, and semantic visual contextual cues for an AVSR task. However, its visual contextual cues are restricted to 3-line math formulas, limiting both the quality and quantity of semantic visual information it can utilize.

\vspace{-1mm}
\section{Chinese-LiPS Dataset}
\vspace{-0.8mm}

\begin{figure*}[t!]
\centerline{\includegraphics[scale=0.49]{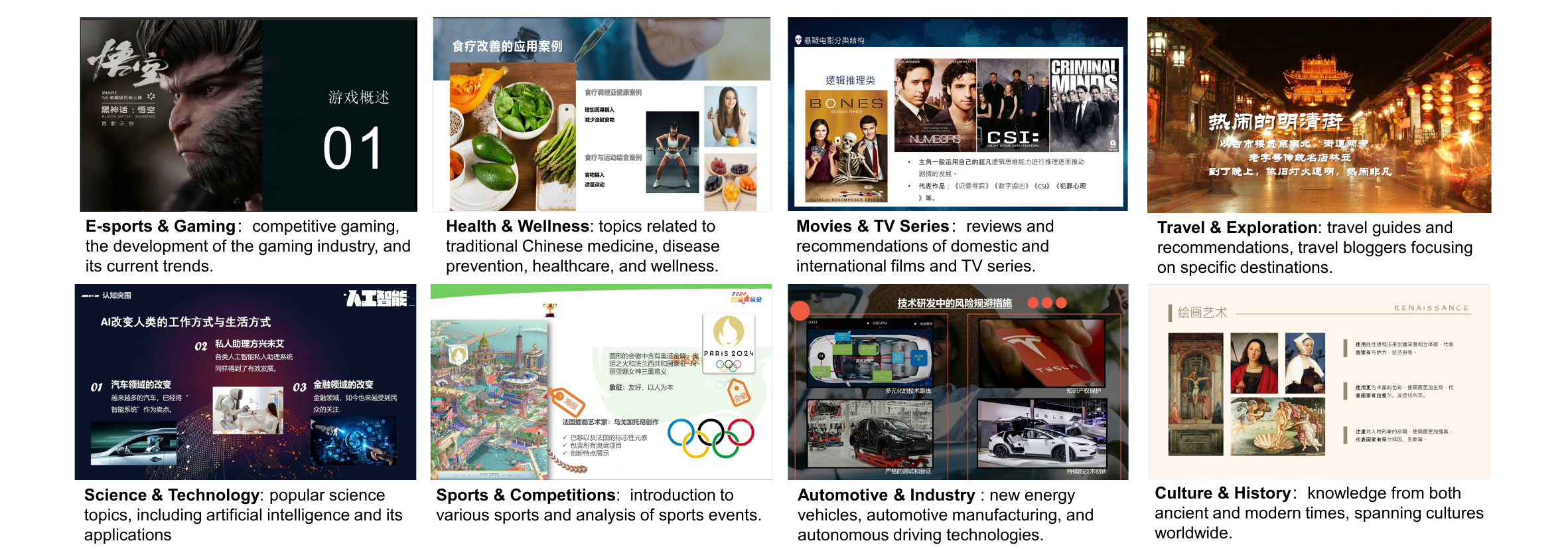}}
\vspace{-2mm}
\caption{Overview of slide styles and themes across different topics in the Chinese-LiPS dataset: we display examples from eight specific topics and the content included in each, while the `Others' topic covers diverse topics such as dance, fashion, cuisine, photography, etc.}
\label{slides}
\vspace{-5.7mm}
\end{figure*}

\subsection{Basic Information}
\vspace{-0.2mm}
The proposed Chinese-LiPS dataset is a multimodal Chinese speech recognition dataset, comprising lip-reading and slide-based videos focused on instruction, lectures, and educational presentations. It contains approximately 100 hours data and 36,208 clips from 207 speakers. The dataset includes speech, slide video, and lip-reading video for each clip. The presentation slides, created by domain experts to ensure content accuracy, are carefully designed to avoid large blocks of text, with speakers delivering content beyond merely reading the slides. Instead, they combine visuals and text effectively, using various examples to ensure the audience can clearly understand the content of the presentation. Each presentation typically includes around 25 slides. The speech data is collected from professional speakers across various fields in China, recorded in quiet, natural environments to ensure high-quality input, with all speakers using Mandarin. 
All components are carefully edited and manually aligned to maintain precision. 
\vspace{-4mm}
\begin{figure}[h!]
\centerline{\includegraphics[scale=0.37]{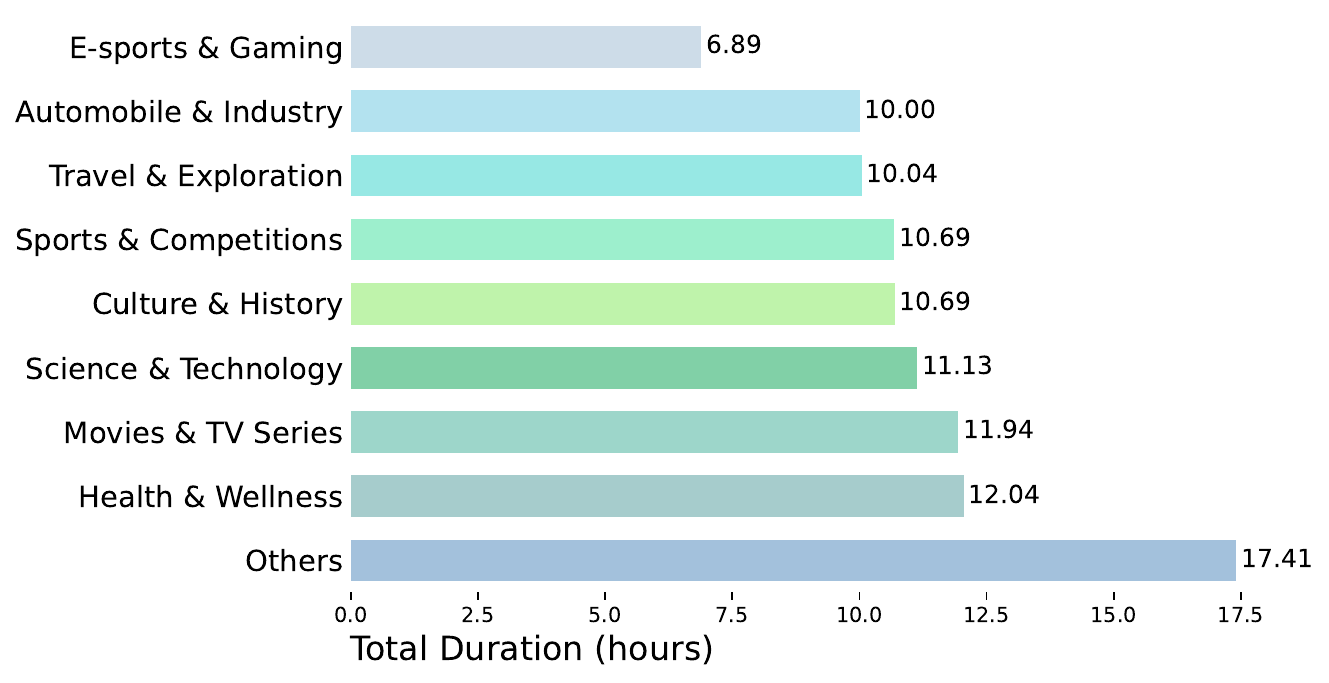}}
\vspace{-1mm}
\caption{Distribution of total recording duration by topic.}
\label{duration}
\end{figure}

\vspace{-8.3mm}
\begin{figure}[h!]
    \centering
    \subfigure[Age distribution.]{\includegraphics[width=0.24\textwidth]{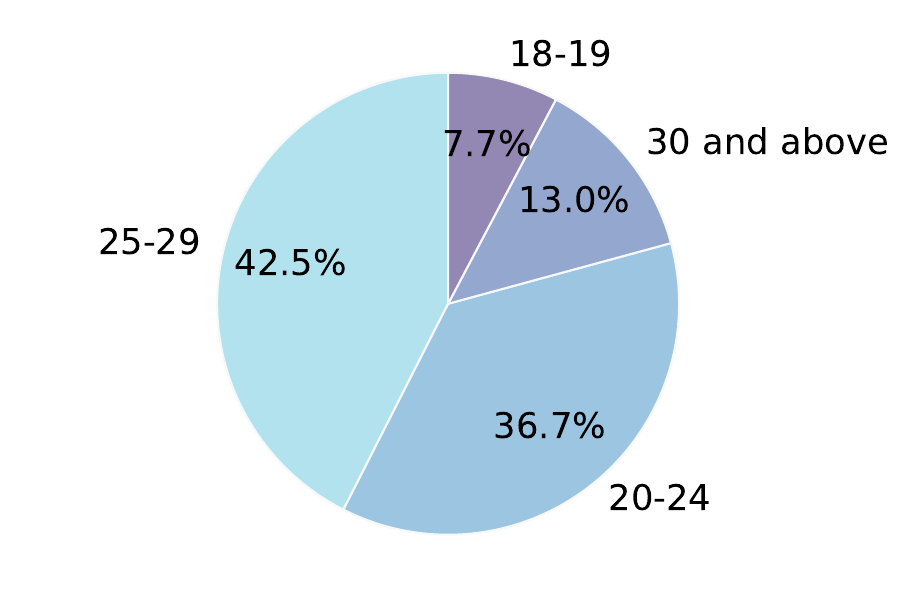}}
    \subfigure[Device distribution.]{\includegraphics[width=0.24\textwidth]{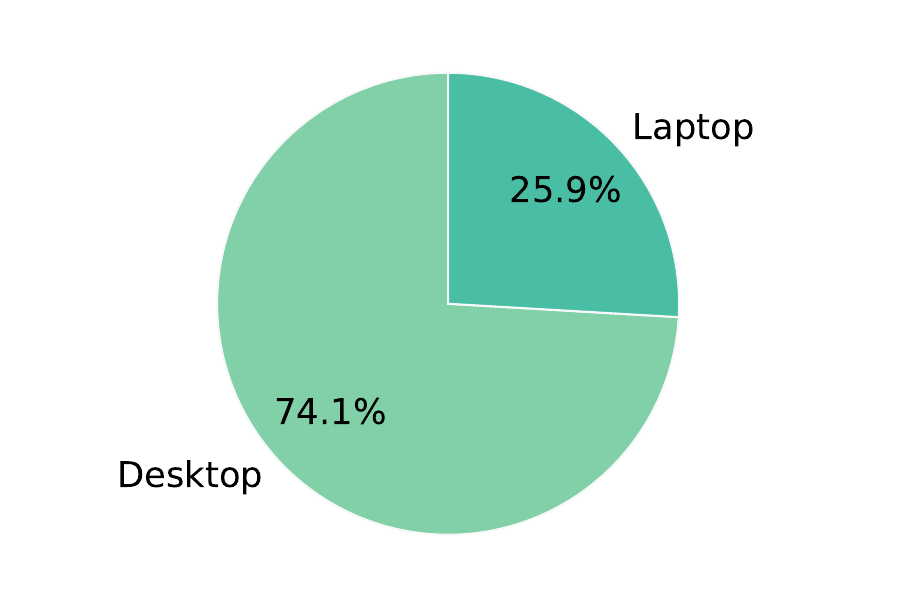}}
    \subfigure[Gender distribution.]{\includegraphics[width=0.24\textwidth]{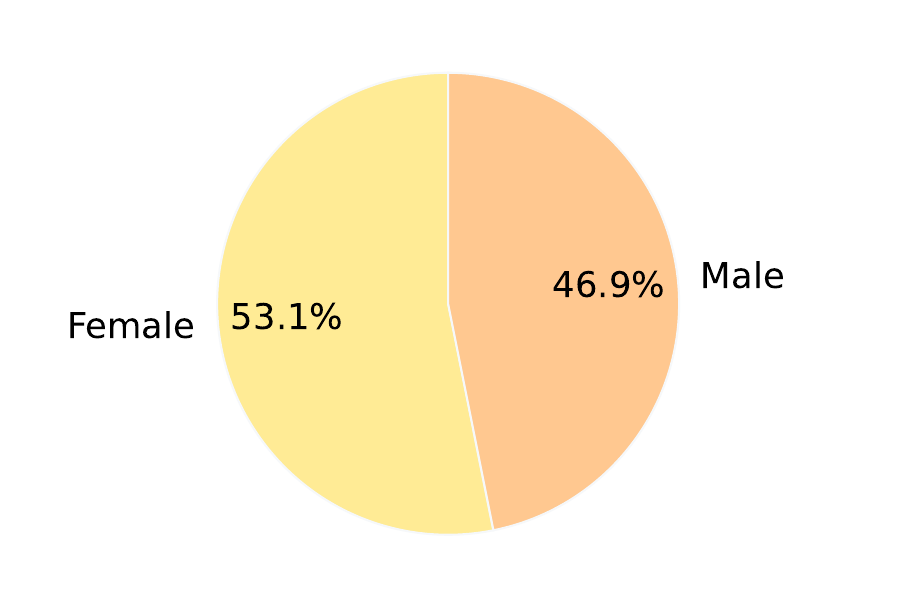}}
    \subfigure[Segment duration.]{\includegraphics[width=0.24\textwidth]{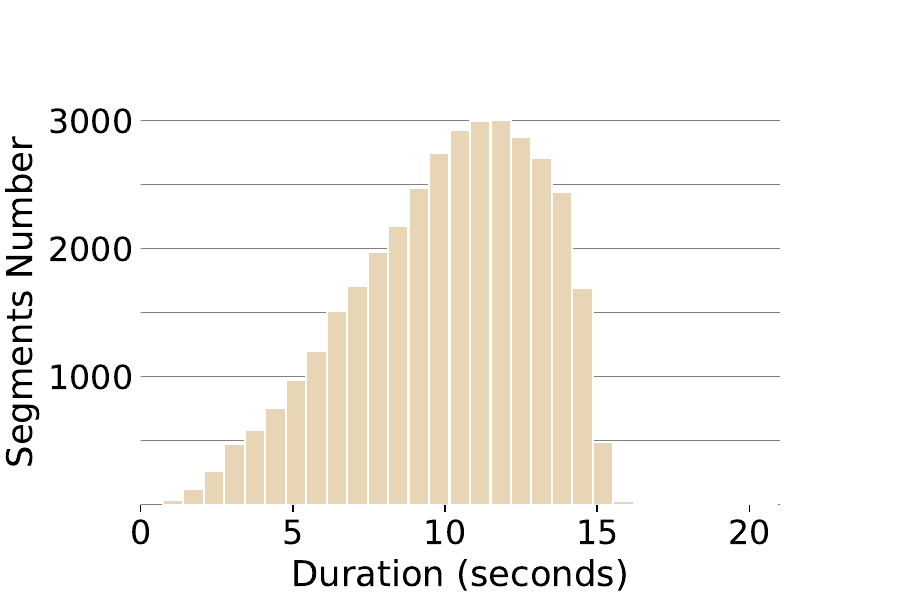}}
    \caption{Distribution analysis of Chinese-LiPS dataset.}
    \label{statistics}
\vspace{-2mm}   
\end{figure}

\subsection{Data Distribution}
\vspace{-0.8mm}

As shown in Fig.~\ref{duration}, the topics of the slides and presentations in our dataset are broadly categorized into nine types: E-sports \& Gaming, Automobile \& Industry, Travel \& Exploration, Sport \& Competitions, Culture \& History, Science \& Technology, Movies \& TV Series, Health \& Wellness, and Others. These topics are highly popular on social media platforms and feature domain-specific terminology.
The distribution of presentation durations across these topics is relatively balanced, ensuring a broad range of instructional content. Fig.~\ref{slides} displays some of our slide examples.

\begin{figure*}[t!]
\centerline{\includegraphics[scale=0.43]{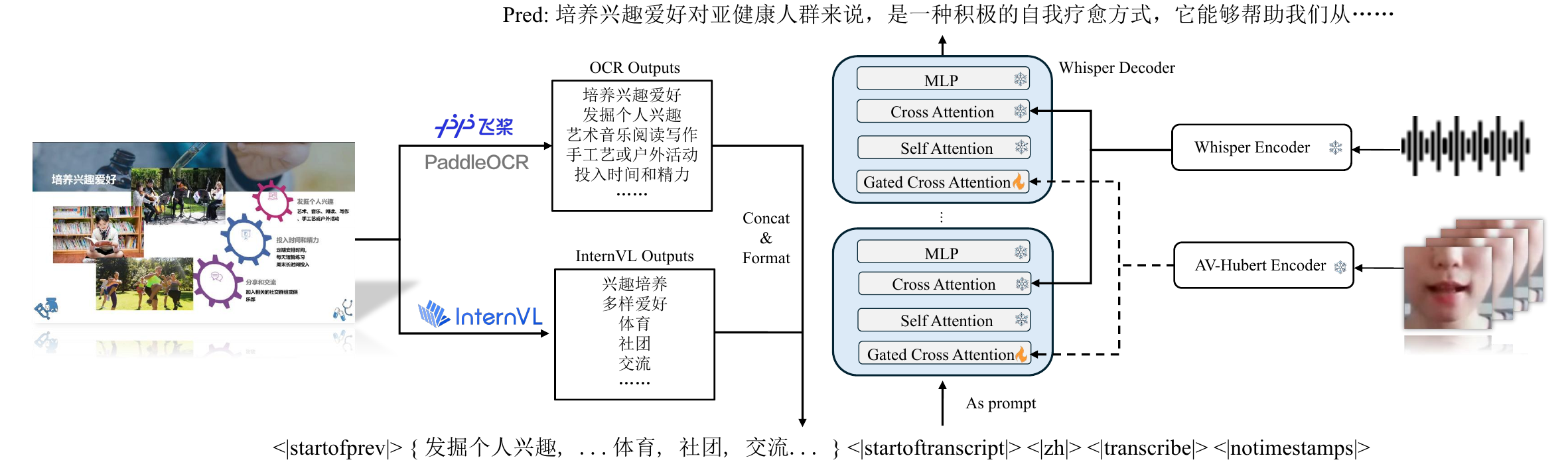}}
\vspace{-2mm}
\caption{LiPS-AVSR pipeline.}
\vspace{-5mm}
\label{pipeline}
\end{figure*}

Fig.~\ref{statistics} provides four key distribution statistics of our dataset: age, devices used, gender and segment duration. For age distribution, the majority of the professionals in relevant fields are between 20 and 30 years old. In terms of device, three-quarters of the recordings are made on desktop computers, predominantly from Dell, while the remaining quarter are recorded on laptops, mainly from Lenovo and Huawei. As for gender, we make efforts to ensure a balanced representation of both male and female speakers to minimize potential gender biases in speech recognition performance. Regarding segment duration, the average speech length is 10 seconds, with all segments not exceeding 30 seconds. 

For the dataset split, we divide Chinese-LiPS into 80\% training, 15\% testing, and 5\% validation sets. To ensure diversity, we maintain a balanced gender and age distribution across the subsets, with both the training and testing sets covering all topics. Additionally, there is no overlap of speakers between different sets, as shown in Table \ref{split_details}.

\begin{table}[b]
\vspace{-8mm}
\caption{split details of Chinese-LiPS Dataset, M:F represents the ratio of male to female speakers}
\vspace{-3mm}
\renewcommand{\arraystretch}{1.2} 
\begin{center}
\setlength{\tabcolsep}{4pt} 
\begin{tabular}{cccccc}
\hline
Split      & Duration(h) & Segments & Speakers & M:F & Topics\\ \hline
Train      & 85.37           & 30341    & 175  &1:1.16 &  9 \\
Test       & 10.12           & 3908     & 21   &1:1.10 &  9 \\ 
Validation & 5.35            & 1959     & 11   &1:0.83 &  6 \\ \hline
All        & 100.84          & 36208    & 207  &1:1.13 &  9 \\ \hline
\end{tabular}
\end{center}
\label{split_details}
\vspace{-4mm}
\end{table}

\vspace{-1.5mm}
\section{Experiments}
\vspace{-0.8mm}

\subsection{Pipeline}
\vspace{-0.8mm}

We conduct experiments using the Whisper and Whisper-Flamingo backbones. Whisper is a Transformer-based end-to-end model that delivers strong performance in speech recognition tasks\cite{whisper}. Whisper-Flamingo extends Whisper by incorporating lip-reading information extracted through AV-Hubert\cite{shilearning} and using a gated cross-attention layer to integrate lip-reading and speech information. This enhancement has achieved state-of-the-art performance on the LRS2 and LRS3 datasets for English speech recognition task\cite{whisperflamingo}.

Fig.~\ref{pipeline} demonstrates the LiPS-AVSR pipeline. First, we use PaddleOCR\footnote{PaddleOCR is an open-source OCR tool developed by PaddlePaddle. Details are here \url{https://github.com/PaddlePaddle/PaddleOCR}.} to extract textual information from slides. To derive semantic information, we utilize InternVL2\cite{internvl2}, a lightweight vision-language pre-trained model that effectively captures semantic content from images and graphical elements. The features extracted by OCR and InternVL2 are presented as Chinese keywords. These keywords are concatenated and integrated into Whisper's prompt, formatted as:
$<|startofprev|>\{text\_prompt\} <|startoftranscript|><|zh|><|transcribe|><|notimestamps|>$,
which serves as input prompt for decoder. 
When the model receives a text prompt during inference, it tends to generate content that aligns with or is related to the prompt, enhancing transcription accuracy\cite{peng2023prompting}. 
Speech features are extracted using the Whisper encoder and fed into the cross-attention layers, while lip-reading features derived from AV-Hubert are provided to the gated cross-attention layers. The decoder then autoregressively generates the predicted text output.

\vspace{-1.5mm}
\subsection{Experments Setup}
\vspace{-0.8mm}

The data preprocessing details are as follows:
\begin{itemize}
\item For the preprocessing of lip-reading and speech data, we adopt the approach described in Auto-AVSR\cite{ma2023auto}. Speech is processed at a sampling rate of 16 kHz, while the lip region in the video is detected and resized to a resolution of 96×96 at 25 frames per second.
\item For presentation slides, we extract the first frame from the slides video and perform OCR. To capture slide semantics, we use the InternVL2 8B model\footnote{\url{https://huggingface.co/OpenGVLab/InternVL2-8B}}, with a prompt to “Summarize the objects in the image using ten words, separated by commas.”
\end{itemize}

We conduct experiments using the Whisper large-v2 model\footnote{\url{https://openai.com/index/whisper/}} to evaluate different input modalities. Specifically, we perform experiments across two main settings:
\begin{itemize}
    \item \textbf{Speech-only}: In this setting, we use the original Whisper model and also fine-tune the entire Whisper model using speech data from the Chinese-LiPS training set. 
    \item \textbf{Speech + Lip-reading}: In this setting, we fine-tune the gated cross-attention layer of the Whisper-Flamingo model using both speech and lip-reading data from the Chinese-LiPS training set.
\end{itemize}

For each setting, we test four different prompt configurations: 1) No prompt. 2) OCR-extracted text only. 3) InternVL2-extracted semantic information only. 4) A combination of OCR and InternVL2 features. These experiments aim to investigate the contributions of various visual and textual inputs to the performance of ASR. For Speech-only setting, we select the model with the best performance as our baseline to ensure a more rigorous evaluation.

\vspace{-1mm}
\subsection{Metric}
\vspace{-0.8mm}

We use Character Error Rate (CER) as the evaluation metric
for our experiments. The formula for calculating CER is:
\vspace{-0.8mm}
\begin{equation}
CER=\frac{S+D+I}{N} \label{eq}
\end{equation}
where S, D, and I represent substitution, deletion, and insertion errors, and N is the total number of characters in the reference.

\vspace{-1mm}
\subsection{Result}
\vspace{-0.8mm}

\begin{figure*}[t]
\centerline{\includegraphics[scale=0.24]{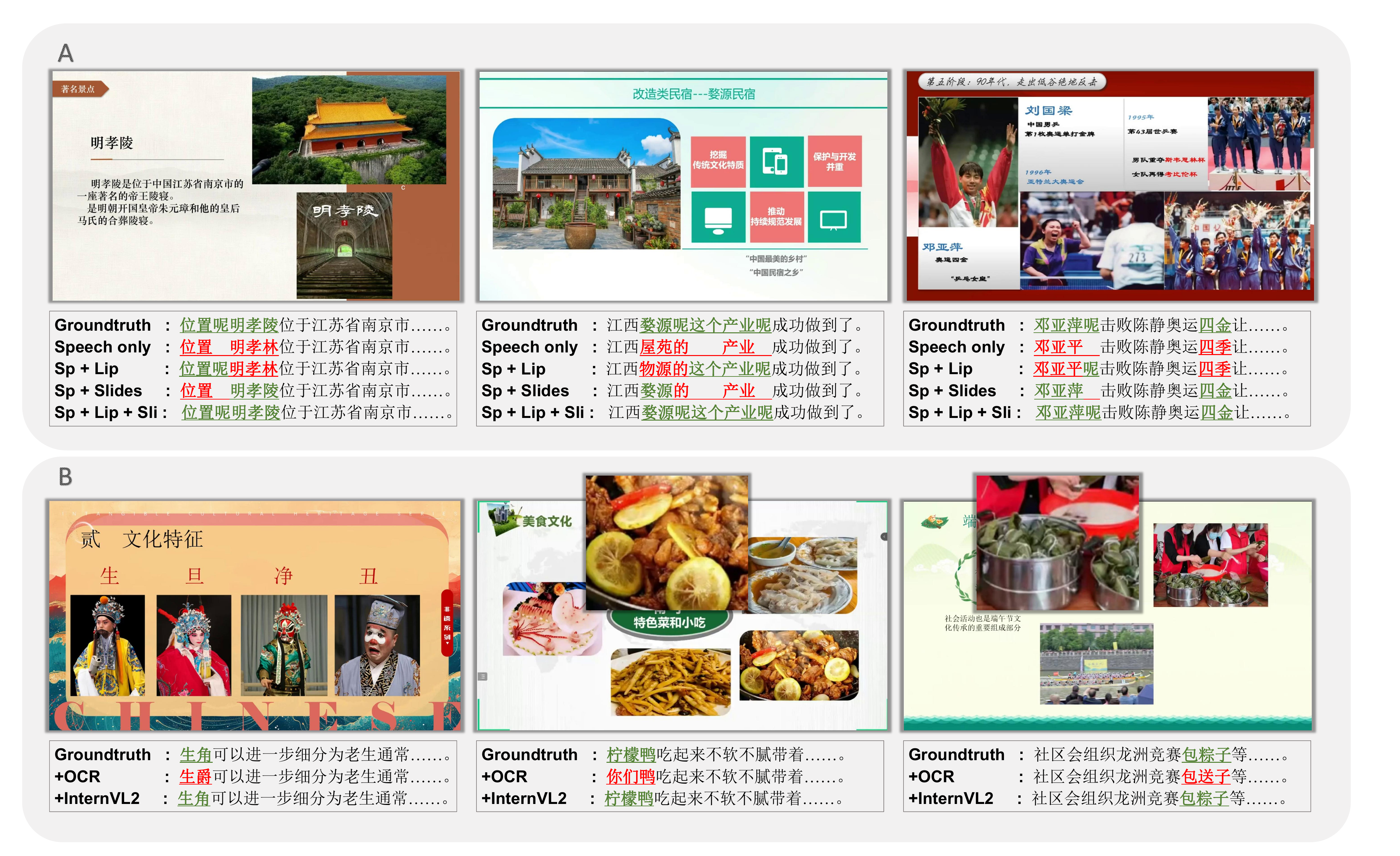}}
\vspace{-2mm}
\caption{Error correction examples using slide and lip-reading information. (A) Lip-reading mitigates hesitation and filler errors, while slide data addresses domain-specific terms. (B) OCR fails to capture visual cues, but InternVL2 effectively extracts meaningful context.}
\label{case}
\vspace{-4mm}
\end{figure*}

As shown in Table \ref{result}, the experimental results highlight the effectiveness of integrating multiple input modalities into the Whisper large-v2 model on the Chinese-LiPS dataset. Using the original Whisper model with only speech achieves a CER of 3.99\%. Incorporating visual information, such as text extracted from slides via OCR and semantic keywords obtained with InternVL2, significantly improves performance. When these two sources are combined, the CER further drops to 2.99\%, representing an approximate 25\% improvement. This demonstrates the complementary strengths of OCR and InternVL2, where their combined information enhances the model's ability to capture meaningful visual context.

Additionally, lip-reading features also contribute substantially to performance gains. When lip-reading data is included and the gated cross-attention layer of the Whisper decoder is fine-tuned, notable improvements are observed. For example, even without other visual cues, ID 5 achieves an 8\% performance improvement over ID 1. Ultimately, combining all visual modalities—lip-reading, OCR-extracted text, and InternVL2-derived semantic information—synergistically lowered the CER to 2.58\%, corresponding to a 35\% improvement. These results emphasize the importance of leveraging multiple complementary sources of information to enhance the robustness and accuracy of speech recognition.

\vspace{-4mm}
\begin{table}[htbp]
\vspace{-2mm}
\caption{Performance Comparison of Whisper-Large-V2 on Chinese-LiPS. The ID represents each modality combination, where \checkmark indicates the inclusion of that modality}
\vspace{-3mm}
\renewcommand{\arraystretch}{1.2} 
\begin{center}
\setlength{\tabcolsep}{4pt} 
\begin{tabular}{cccccc}
\hline
ID          & Speech & Lip-reading & OCR & InternVL2 & CER(\%)$\downarrow$ \\ \hline
1(baseline) & \checkmark      &             &     &     & 3.99   \\
2           & \checkmark      &             & \checkmark   &     &3.37 \\
3           & \checkmark      &             &     & \checkmark   & 3.33 \\
4           & \checkmark      &             & \checkmark   & \checkmark   & 2.99  \\ \hline
5           & \checkmark      & \checkmark           &     &     & 3.69  \\
6           & \checkmark      & \checkmark           & \checkmark   &     & 2.74 \\
7           & \checkmark      & \checkmark           &     & \checkmark   & 3.05   \\
8           & \checkmark      & \checkmark           & \checkmark   & \checkmark   & \textbf{2.58} \\ \hline
\end{tabular}
\end{center}
\label{result}
\vspace{-7mm}
\end{table}

\vspace{-1.5mm}
\subsection{Analysis}
\vspace{-0.8mm}
Fig.~\ref{case} consists of two parts that illustrate the effectiveness of combining slide and lip-reading information in addressing substitution and deletion errors. In Fig.~\ref{case} (A), we showcase specific instances where integrating these modalities yields significant error corrections. For example, adding lip-reading information helps recover filler words, while incorporating presentation slide data improves recognition of rare terms such as geographical place names. Combining both modalities ensures that all errors are addressed comprehensively. Fig.~\ref{case} (B) highlights examples where OCR alone fails to capture semantic information from visual cues, but the InternVL2 effectively resolves these gaps. For instance, in the middle image, the slide features `lemon duck', yet OCR fails because the word `lemon' is not present. InternVL2, however, identifies `lemon' as one of the ten summarized keywords extracted from the image, enabling accurate correction.

These examples mirror real-world lecture scenarios, where slides provide crucial domain-specific terms, and lip movements enhance the understanding of speech variations. Together, these modalities create a robust complementarity for achieving high ASR performance in complex instructional settings.

\begin{table}[t]
\caption{Error Analysis Across Different Modalities in the Test Set: Total Number of Chinese Characters = 150,059}
\vspace{-3mm}
\begin{center}
\renewcommand{\arraystretch}{1.2} 
\setlength{\tabcolsep}{4pt} 
\begin{tabular}{ccccc}
\hline
ID      & Modality & Substitution$\downarrow$ & Deletion$\downarrow$ &Insertion$\downarrow$ \\ \hline
1      &Speech only &3851 &1697 &437     \\
4      &Speech + Slides&3531$\downarrow$ &447$\downarrow$ &510 \\
5      &Speech + Lip &4499 &509$\downarrow$ &522    \\
8      &Speech + Lip + Sli  &3047$\downarrow$ &335$\downarrow$ &484   \\ \hline
\end{tabular}
\end{center}
\label{SDI}
\vspace{-7mm}
\end{table}

As shown in Table \ref{SDI}, the error analysis across different modalities reveals distinct contributions from lip-reading and slide information. Lip-reading primarily reduces deletion errors, decreasing them from 1697 to 509 in ID 5. This improvement aligns with our hypothesis that lip-reading conveys articulation-related information, helping to recover elements often missing in speech, such as filler words, hesitation markers, and incomplete speech segments. In contrast, presentation slides contribute significantly to reducing substitution errors, dropping from 3851 to 3531 in ID 4. They also help address some deletion errors, especially for key domain-specific terms. These results align with our hypothesis that slides provide semantic and contextual information that is crucial for recognizing specialized vocabulary, proper nouns, and homophones. 

The combination of lip-reading and slide information highlights their complementary nature: while lip-reading excels in addressing articulation-related issues, slides capture textual and contextual details, together forming a robust system that improves overall speech recognition performance in complex instructional environments.

\vspace{-1.0mm}
\section{Conclusion}
\vspace{-0.8mm}
In this paper, we introduce Chinese-LiPS, the first high-quality multimodal Chinese AVSR dataset integrating both lip-reading and presentation slide information. Based on this dataset, we propose a simple yet effective pipeline, LiPS-AVSR, which leverages lip-reading, OCR-extracted text, and semantic information from slides to significantly enhance speech recognition performance and validate the complementary roles of these multiple visual modality.
\bibliographystyle{IEEEbib}
\normalem
\bibliography{main}

\end{document}